\documentclass[iop,apj,12pt,numberedappendix]{emulateapj}
\usepackage{times}
\usepackage[normalem]{ulem}
\usepackage{graphicx}
\usepackage{natbib}
\usepackage{amsfonts}
\usepackage{amsmath}
\usepackage{multirow}
\usepackage{enumerate}
\usepackage{comment}
\usepackage{float}
\usepackage{verbatim}
\usepackage[usenames,dvipsnames]{xcolor}
\usepackage[plainpages=false, colorlinks=true, anchorcolor=blue, linkcolor=blue, citecolor=blue, bookmarks=false]{hyperref}
\newcommand{\be}{\begin{eqnarray}}
\newcommand{\ee}{\end{eqnarray}}

\newcommand{\shortauth}{Morozova et al.}
\newcommand{\slugcom}{Submitted to ApJ, 03/28/2016}
\slugcomment{\slugcom}

\lefthead{\sc \footnotesize \slugcom \hfill \shortauth}
\righthead{\sc \footnotesize \slugcom \hfill \shortauth}

\begin{document}

\title{Numerical Modeling of the Early Light Curves of Type IIP Supernovae}

\author{Viktoriya Morozova\altaffilmark{1}}
\author{Anthony L. Piro\altaffilmark{2}}
\author{Mathieu Renzo\altaffilmark{3,1}}
\author{Christian D. Ott\altaffilmark{4,1}}
\altaffiltext{1}{TAPIR, Walter Burke Institute for Theoretical Physics, MC 350-17,
  California Institute of Technology, Pasadena, CA 91125, USA, 
  morozvs@tapir.caltech.edu}
\altaffiltext{2}{Carnegie Observatories, 813 Santa Barbara Street, Pasadena, CA 91101, USA}
\altaffiltext{3}{Anton Pannekoek Astronomical Institute, University of Amsterdam, 1090 GE Amsterdam, The Netherlands}
\altaffiltext{4}{Yukawa Institute for Theoretical Physics, Kyoto University, Kyoto, Japan}

\begin{abstract}
The early rise of Type IIP supernovae (SN IIP) provides
important information for constraining the properties of their
progenitors.  This can in turn be compared to pre-explosion imaging
constraints and stellar models to develop a more complete picture of
how massive stars evolve and end their lives.  Using the SuperNova
Explosion Code (\texttt{SNEC}), we model the first $40$ days of SNe
IIP to better understand what constraints can be derived from their
early light curves.  We use two sets of red supergiant progenitor
models with zero-age main sequence masses in the range between
$9\,M_{\odot}$ and $20\,M_{\odot}$. We find that the early properties
of the light curve depend most sensitively on the radius of the
progenitor, and thus provide a relation between the $g$-band rise time
and the radius at the time of explosion. This relation will be useful
for deriving constraints on progenitors from future observations,
especially in cases where detailed modeling of the entire rise is not
practical.  When comparing to observed rise times, the radii we find
are a factor of a few larger than previous semi-analytic derivations
and generally in better agreement with what is found with current
stellar evolution calculations.
\end{abstract}

	
  
\section{Introduction}

One of the outstanding problems in astrophysics is connecting the
variety of core-collapse supernovae (SNe) we observe
with the massive progenitors that give rise to them. Ideally, we would
use pre-explosion imaging to directly identify these progenitors
\citep[e.g.,][]{vandyk:12,smartt:09,li:06}. Unfortunately, in most
cases such information is not available because the progenitor is too
dim or deep pre-explosion imaging is not available at the needed
location. For this reason, features of the early light curves can be
especially helpful in constraining progenitor properties
\citep{piro:13}. Emission dominated by shock cooling should reflect
the radius of the exploding star (\citealt{nakar:10}, hereafter
NS10). Furthermore, the presence of extended material
\citep{nakar:14,piro:15} and the interaction with a companion
\citep{kasen:10} can create additional features that teach us about
how massive stars end their lives.

This early phase has been explored in a number of theoretical works
semi-analytically (e.g., NS10; \citealt{piro:10,rabinak:11}).  These
studies generally took the approach of assuming an idealized (i.e.,
polytropic) density profile for the star to make
connections between early light curve properties and the properties of
the star in a general way. The question though is whether in nature
such idealized profiles actually occur at relevant depths within the
star.  On the other hand, numerical models represent a powerful
complementary approach to analytic studies of SN light curves,
including the early phases
\citep[see][]{eastman:94,young:04,kasen:09,bersten:11,dessart:13}.
For example, \citet{bersten:12} constrained the radius of the
progenitor star for SN 2011dh based on the numerical modeling of its
early light curve.  In this case, many of these calculations are
tailored for specific events. This makes it difficult to generalize
these results more broadly, which would be especially useful as
current and future transient surveys (e.g., iPTF/ZTF, ASAS-SN,
BlackGEM, Pan-STARRS, LSST) find larger samples of SNe.

Motivated by these issues, we undertake a numerical exploration of how
rise times of SNe IIP vary with the parameters of their
progenitors. We utilize two sets of nonrotating solar-metallicity
progenitor models from the stellar evolution codes \texttt{MESA}
\citep{paxton:11,paxton:13,paxton:15} and \texttt{KEPLER}
\citep{weaver:78,woosley:07,sukhbold:14,woosley:15}, and explode these
models and generate light curves using the SuperNova Explosion
Code (\texttt{SNEC}; \citealt{morozova:15}). We describe important
differences between realistic stellar models and more idealized
treatments, and how these impact observable features of the early
light curves. In addition, based on the results of our numerical
models, we derive a relation for the rise time $t_{\rm rise}$ of
SN IIP light curves in $g$-band as a function of the radius $R$ of the
progenitor. This relation may be useful in the future analyses of
observed early rises, especially when comparing large samples of
events.

The paper is organized as follows. In Section~\ref{sec0}, we describe
the progenitor models used in this study and the numerical setup of
our simulations. In Section~\ref{sec1}, we discuss two important
aspects of our calculations related to the stellar models and
radiative diffusion. This highlights differences with more idealized
treatments. In Section~\ref{sec2}, we present our full set of
explosion models and summarize how the properties of the
light curve rise relate to the radius and ejecta
mass.  In Section~\ref{sec3}, we summarize our main results and
discuss future work.


\section{Numerical setup}
\label{sec0}

This work is primarily focused on SNe IIP, and thus we only consider
red supergiant (RSG) progenitor models since these are the
observationally-confirmed SN IIP progenitors \citep[see
  e.g.,][]{smartt:09,fraser:12,maund:13}.  We consider two sets of
nonrotating solar-metallicity pre-collapse RSG models. The first set
of models comes from the stellar evolution code \texttt{KEPLER}
\citep{woosley:07,woosley:15,sukhbold:14,sukhbold:16} and has zero-age
main sequence (ZAMS) masses in the range between $9\,M_{\odot}$ and
$20\,M_{\odot}$ in steps of $0.5\,M_{\odot}$.  We refer the reader to
\citet{sukhbold:16} for a detailed description of these models.  The
second set of the models we generate\footnote{We provide full details
  to reproduce the \texttt{MESA} models at
  \url{https://stellarcollapse.org/Morozova2016}. There, we also
  provide the lightcurves resulting from our model calculations.}
using the stellar evolution code \texttt{MESA}
\citep{paxton:11,paxton:13,paxton:15}, revision 7624. These models
have ZAMS masses in the range between $11\,M_{\odot}$ and
$20\,M_{\odot}$ in steps of $0.5\,M_{\odot}$. Our calculation use the
same parameter set described in \cite{morozova:15}. We summarize for
completeness the parameters that influence more directly the radius
determination. We use the 21-isotope nuclear reaction network
\texttt{approx21.net}, and use the Ledoux criterion for convection,
following \cite{sukhbold:14} for the choice of the free
parameters. This corresponds to a mixing length parameter
$\alpha_\mathrm{mlt}=2.0$, exponentially decreasing overshooting (both
for the core and the convective shells) with $f_\mathrm{ov}=0.025$ and
$f_0=0.05$ \citep[see ][]{paxton:11}, and semiconvection efficiency
$\alpha_\mathrm{sc}=0.1$. Wind mass loss is included as in
\cite{morozova:15}, using the ``Dutch'' mass loss algorithm without
any modifying efficiency factor. We set \texttt{mesh\_delta\_coeff} =
\texttt{mesh\_delta\_coeff\_for\_highT} = 1.0 for the spatial
resolution, and \texttt{varcontrol\_target} = $10^{-4}$ for the
temporal resolution. We note that experiments with \texttt{MESA} show
that the pre-collapse RSG radius depends sensitively on wind
efficiency (Renzo et al., in preparation), overshooting, and
mixing-length parameters. Generally speaking, the higher the wind
efficiency, the more mass is removed and the smaller the radius.  The
larger $f_\mathrm{ov}$, the more massive and luminuous the He core and
the larger the radius. The larger $\alpha_\mathrm{mlt}$, the more
efficient the energy transport through the envelope and the smaller
the radius (cf.\ also \citealt{dessart:13}). Ultimately, 3D
radiation-hydrodynamic models will be needed to robustly predict RSG
radii, e.g., \cite{chiavassa:09}.

\begin{figure}[t]
  \centering
  \includegraphics[width=0.475\textwidth]{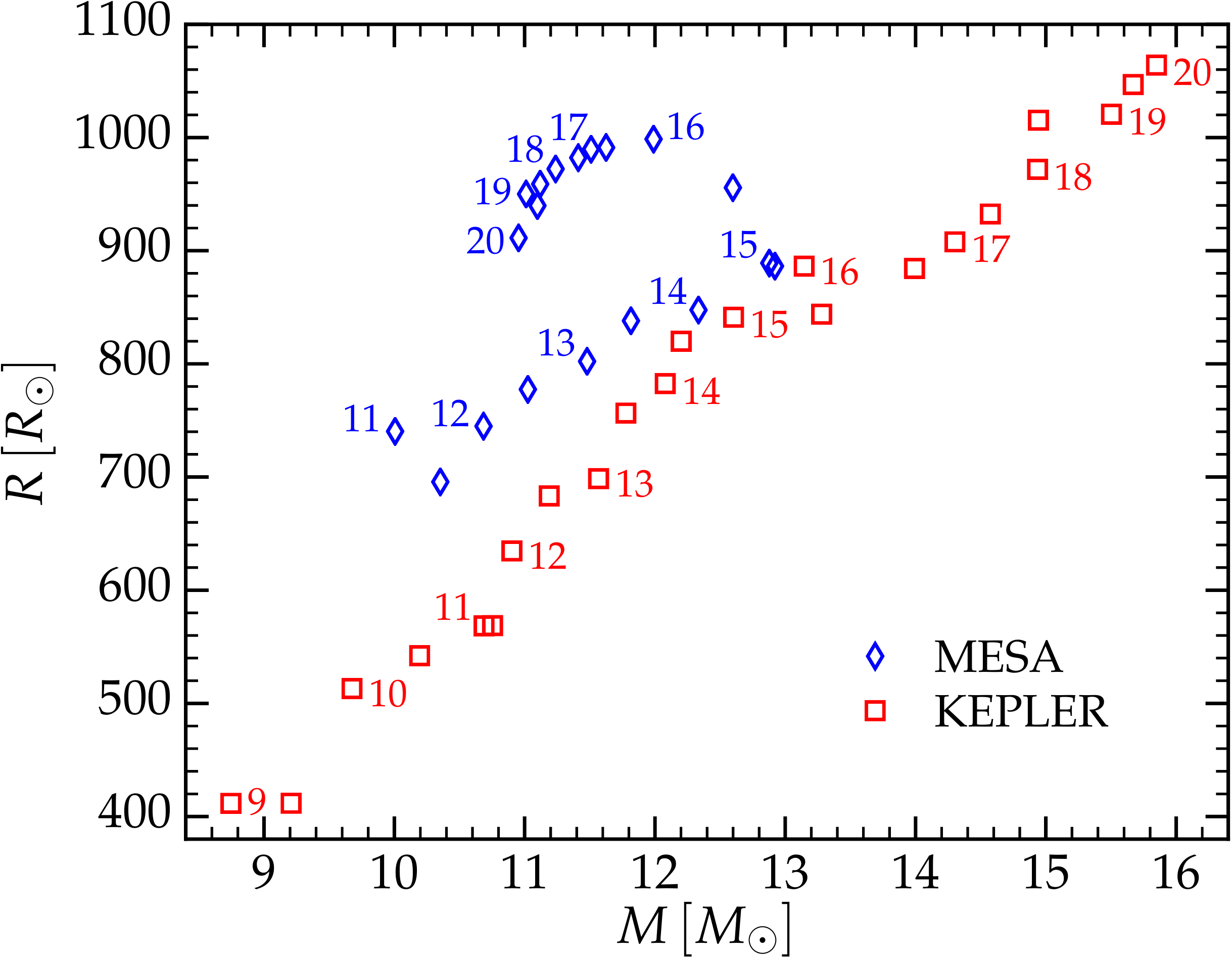}
  \caption{Radii of the progenitor models versus total masses at the
    onset of core collapse for both \texttt{KEPLER} and \texttt{MESA}
    sets. The numbers give the ZAMS masses of the
    models in units of $M_{\odot}$. In-between models are not labeled
    to avoid clutter.} \label{fig:models}
\end{figure}

Figure~\ref{fig:models} summarizes the main features of
 the considered progenitor models, the
radii and masses at the onset of core collapse. The \texttt{KEPLER} models show
a strong correlation between radius and mass. We emphasize that
whether or not one set of models is more ``correct'' (in the sense
that it is a closer representation of what actually occurs in nature)
is irrelevant to our work.  The reason is that we are looking for
general trends that are satisfied by the explosion of any stellar
model (as we will summarize in Section~\ref{sec2}).  The fact that so
much diversity is seen in Figure~\ref{fig:models} is actually a strength and
not a weakness for our study.
 
All the stellar models are exploded with \texttt{SNEC},
which is described in detail in \cite{morozova:15}. 
We excise the inner $1.4\,M_{\odot}$ of the models,
assuming that this part collapses and forms a neutron star.  Later in
the paper, the ejecta mass, $M_{\rm ej}$, is defined as the total
stellar mass at core collapse minus $1.4\,M_{\odot}$. We
do not model the fallback of material
onto the remnant.  For the current study, we use a thermal bomb
mechanism for the explosion with a duration of $0.001\,{\rm s}$ and a
spread of $0.02\,M_{\odot}$ (the choice of bomb duration is
discussed in Section~\ref{sec2}).  We define the final energy, $E_{\rm
  fin}$, as the final (asymptotic) explosion energy of the
  model. Note that $E_\mathrm{fin}$ is not equal to the energy of the
  thermal bomb that we use to initiate explosion.  The latter is equal
  to the difference between $E_{\rm fin}$ and the total (mostly
  gravitational) energy of the progenitor before
  explosion.  We point out that the
absolute value of the initial gravitational energy of our models can
be of the same order or even larger than $E_{\rm fin}$.  ``Boxcar''
smoothing of the compositional profiles is performed as in
\citet{morozova:15} and the same values for the opacity floor are
used.  We do not include radioactive $^{56}{\rm Ni}$ in our models,
since it does not impact these early phases.  The numerical gridding
of each model is identical to that
used in \citet{morozova:15} and consists of $1000$ cells in mass
coordinate.  However, for a special case
described in Section~\ref{sec1}, we use a grid consisting of $2000$
cells, with increasing resolution toward the surface and the center of
the models.  All light curves are generated for the first $40$ days
after explosion.

\section{Key factors determining early light curves}
\label{sec1}

Before presenting our full set of explosion calculations, it is
helpful to discuss some of the key aspects of the stellar models and
the properties of the radiative diffusion that determine the rising
light curves we find. The two main factors we focus on are the shallow
density profile of RSGs and the time-dependent position of the
so-called luminosity shell. The location of the luminosity shell is
the depth from which photons diffuse to reach the photosphere at a given
time after shock breakout.  These two issues are very connected, since
the density profile determines how the shock accelerates and will
eventually set the depth of the luminosity shell as shock cooling sets
in.

\subsection{Density profiles of RSGs}
\label{profiles}

\begin{figure}[t]
  \centering
  \includegraphics[width=0.475\textwidth]{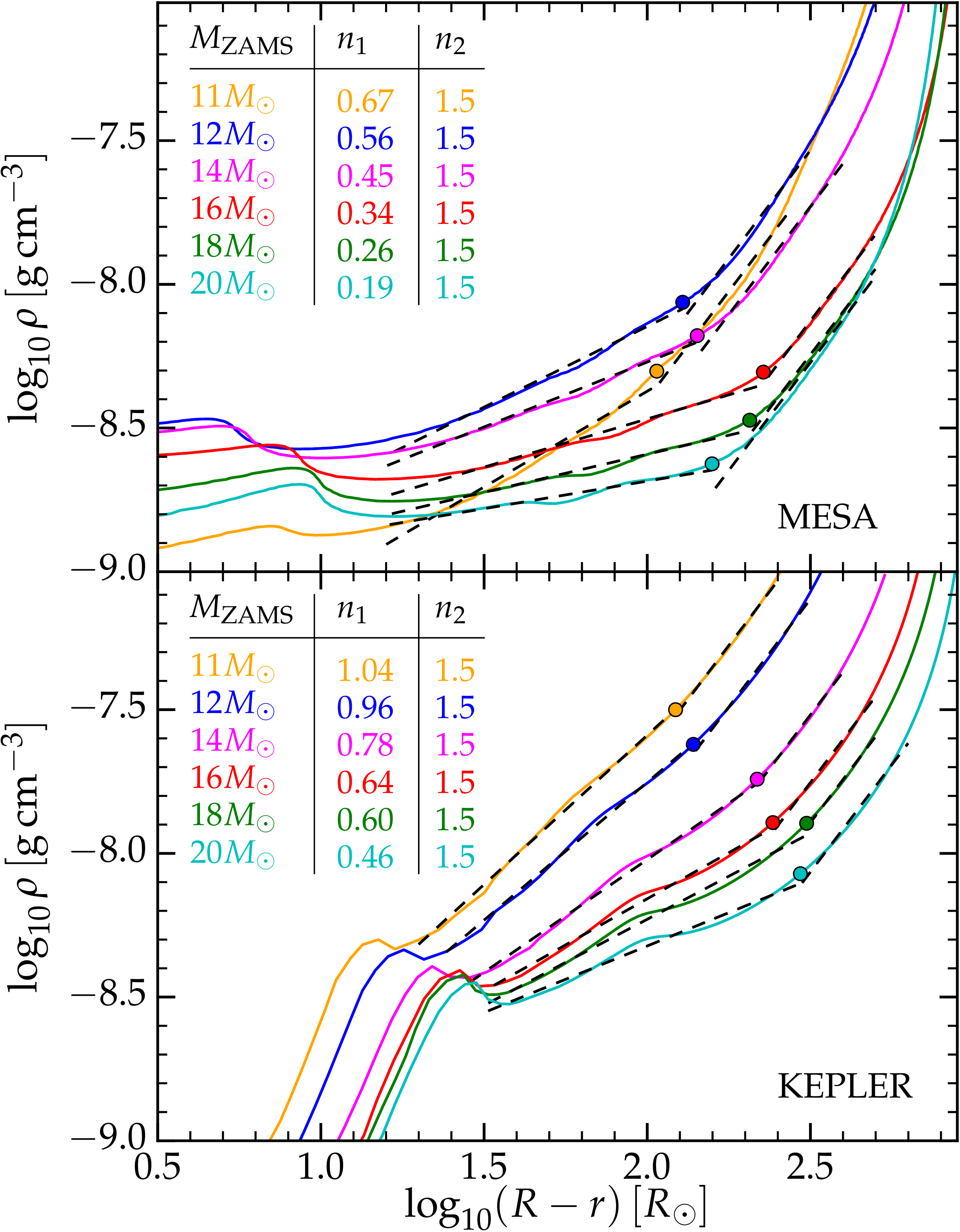}
  \caption{Example density profiles for some of the considered
    pre-collapse RSG models.  Black dashed lines show the
    $\rho\propto(R-r)^n$ ($R$ is the stellar radius
    and $r$ is the distance of the location from the center of the star)
    fits to different regions of the profiles. The variables $n_1$ and
    $n_2$ are the power-law exponents of the outer and inner parts of
    the profiles, respectively. Colored circles separate the two
    regions with different power-law exponents.} \label{fig:density}
\end{figure}

RSGs have convective envelopes, which means that nominally their
density $\rho$ obeys the power-law $\rho\propto(R-r)^n$, where $r$ is
the distance from the center of the star, $R$ is the radius of the
star, and $n=1.5$.  For this reason, previous analytic studies paid
special attention to density profiles with $n=1.5$
\citep[see][NS10]{matzner:99,rabinak:11}. To test this assumption, we plot in
Figure~\ref{fig:density} $\log_{10}\rho$ as a function of
$\log_{10}(R-r)$ for a representative subset of the models described
in Section~\ref{sec0}. The rest of the models have a similar structure
 and they are not
plotted for clarity. Black dashed lines show fits of the power-laws
$\rho\propto(R-r)^n$ to different regions
of each profile. The indices $n_1$ and
$n_2$ give the power-law exponents obtained for the outer and inner
parts of the profile, respectively. The innermost and outermost
boundary of the fits for each model are chosen in such a way, that the
inner power-law exponent is $1.5$ for
all models since the bulk of the envelope is clearly convective.

From the comparison shown in Figure~\ref{fig:density}, there are
multiple key points to take
away. First, indeed at sufficiently large depths within the envelope
the density profile obeys an $n=1.5$ polytrope. Second, in both 
\texttt{MESA}
and \texttt{KEPLER} models, the outer density profile is
different from $n=1.5$ and in general shallower (although at
sufficiently short distances to the surface the \texttt{KEPLER} models are
steeper, these regions do not impact the light curves we study). This
region can cover a few hundred solar radii. Whether or not the shallow
profile region has an important impact on the rise depends on the
depth where photons are diffusing from during the shock cooling
phase. We address this next.

\begin{figure}[t]
  \centering
  \includegraphics[width=0.475\textwidth]{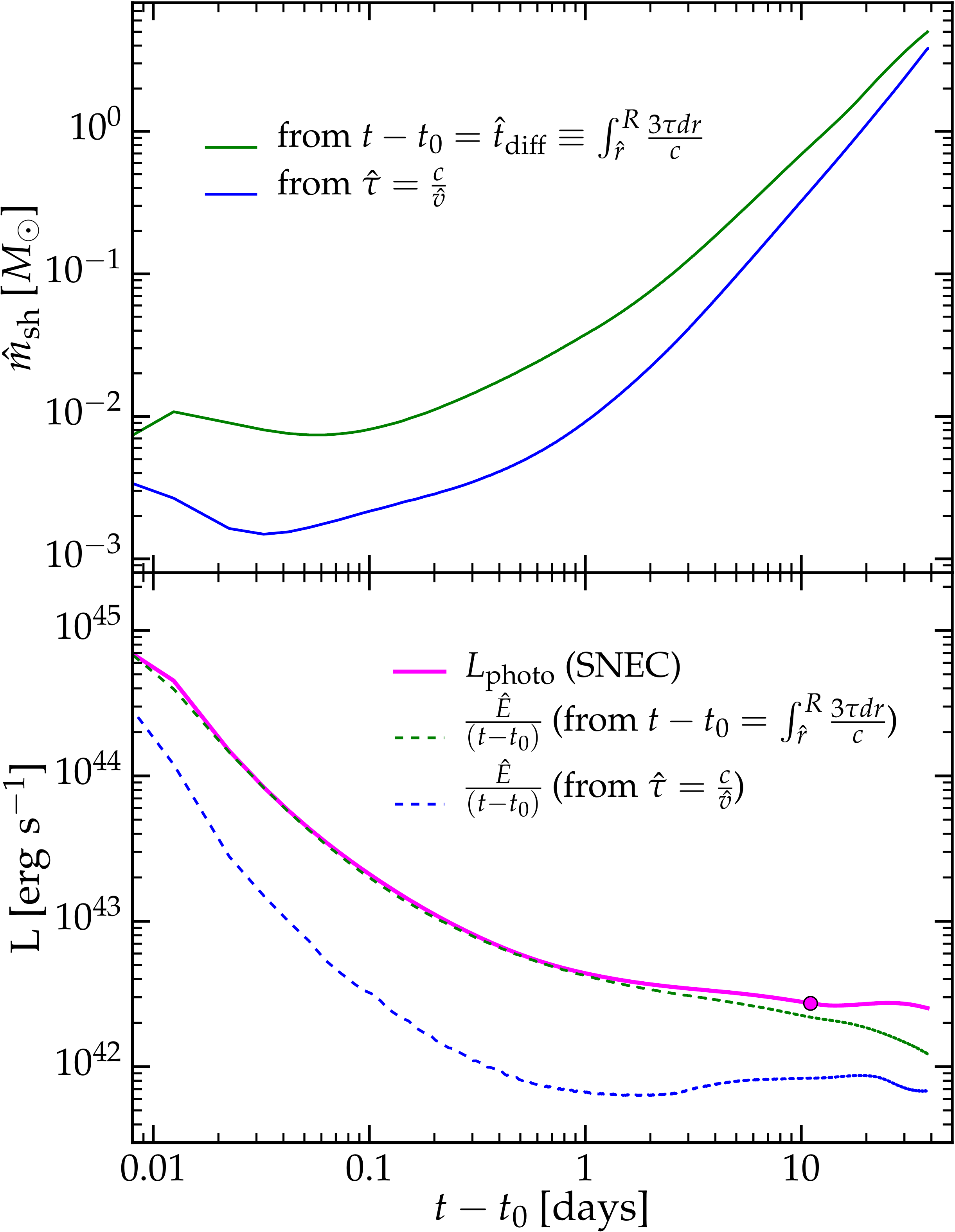}
  \caption{Top panel: $\hat{m}_{\rm sh}$, the difference between the
    total mass of the model and the mass coordinate of the luminosity
    shell, as a function of time since shock breakout ($t_0$) for a
    \texttt{MESA} progenitor model with $M_{\rm ZAMS}=14\,M_{\odot}$
    and $E_{\rm fin} = 10^{51}\,{\rm erg}$. $\hat{m}_\mathrm{sh}$ is
    computed either via the detailed diffusion time estimate in
    Equation~(\ref{tdiff}) or via the simpler estimate in
    Equation~(\ref{eq:approxtdiff}), using $\hat{\tau} = c /
    \hat{v}$. Bottom panel: Bolometric luminosities of the same model,
    computed in three different ways, as described in
    Section~\ref{shells}.  The colored circle indicates the time when
    the luminosity shell found via $\hat t_{\rm diff} = t - t_0$
    enters the convective ($n=1.5$) region of the envelope. Note that
    the \texttt{SNEC} run used to create this figure was carried out
    with 2000 cells in mass coordinate.} \label{fig2}
\end{figure}

\subsection{Position of the luminosity shell}
\label{shells}

The shock-cooling phase (from a few to $\sim20$ days) and the plateau
phase (from $\sim20$ to $\sim100$ days) of SNe IIP light curves are
powered by the energy of the post-explosion shock wave.  This energy
diffuses out of the expanding envelope and at each moment of time we
see the photons coming from the luminosity shell.  The position of the
luminosity shell at time $t$ is
defined by the condition $\hat t_{\rm diff}
= t - t_0$, where $t_{0}$ is the time of shock breakout, $t_{\rm
  diff}=t_{\rm diff}(r,t)$ is the diffusion time at each
 time at each depth, and the hat
indicates the value of this quantity taken specifically at the
luminosity shell. Once the position of the luminosity shell is found,
the observed bolometric luminosity can be estimated as
$\hat{E}/(t-t_{0})$, where $\hat{E}$ is the internal energy of the
luminosity shell.

Given the relatively shallow outer-envelope density profiles we find in realistic
stellar models, we explore where the luminosity depth is at each time.
In integral form, the diffusion time at the luminosity shell is
\begin{equation}
\label{tdiff}
\hat t_{\rm diff} = \int_{\hat{r}}^{R} \frac{3 \tau dr}{c}\ ,
\end{equation}
where
\begin{equation}
\label{tau}
\tau (r) = \int_{r}^{R}\kappa\rho dr\ .
\end{equation}
The top panel of Figure~\ref{fig2} shows the position
(in mass coordinate) of the luminosity shell found in
our calculations for the
\texttt{MESA} model with $M_{\rm ZAMS} = 14\,M_{\odot}$ and
final energy $E_{\rm fin}= 10^{51}\,{\rm
  erg}$.  The radius of this model is $R=848\,R_{\odot}$ and the
ejecta mass is $M_{\rm ej} = 10.93\,M_{\odot}$.  The plotted quantity
$\hat{m}_{\rm sh}$ is the difference between the total mass of the
model and the mass coordinate of the luminosity shell as a function of
time. The green curve finds $\hat{m}_{\rm sh}$ using our integral
definition for $\hat t_{\rm diff}$ given in Equation~(\ref{tdiff}). From
this, one can see where photons are diffusing from at
 a given time.  The blue curve
finds $\hat{m}_{\rm sh}$ using the condition $\hat{\tau} = c/\hat{v}$
for defining the luminosity shell, where $v$ is the velocity of
matter.  This alternative relation (in comparison with the integral we
use in Equation \ref{tdiff}) arises from simplifying
the integral definition for the diffusion time by taking
\begin{equation}
t_{\rm diff} \approx \frac{\hat{\tau}\hat{d}}{c}\ ,
\label{eq:approxtdiff}
\end{equation} 
where $\hat{d}$ is the width of the luminosity shell, and then setting
$\hat{d}=\hat{v}(t-t_0)$. This simpler description is often used in
analytic and semi-analytic works (NS10; \citealt{piro:12}, and
references therein). In general, the integrated diffusion depth is
larger at any given time. This is because the relatively shallow
density profile of realistic models makes the
  full integral crucial for deriving the optical depth.  In contrast,
if the density profile was steeper, only the conditions at the
luminosity depth would really be important and the integral would not
be as important.

\begin{figure}[t]
  \centering
  \includegraphics[width=0.482\textwidth]{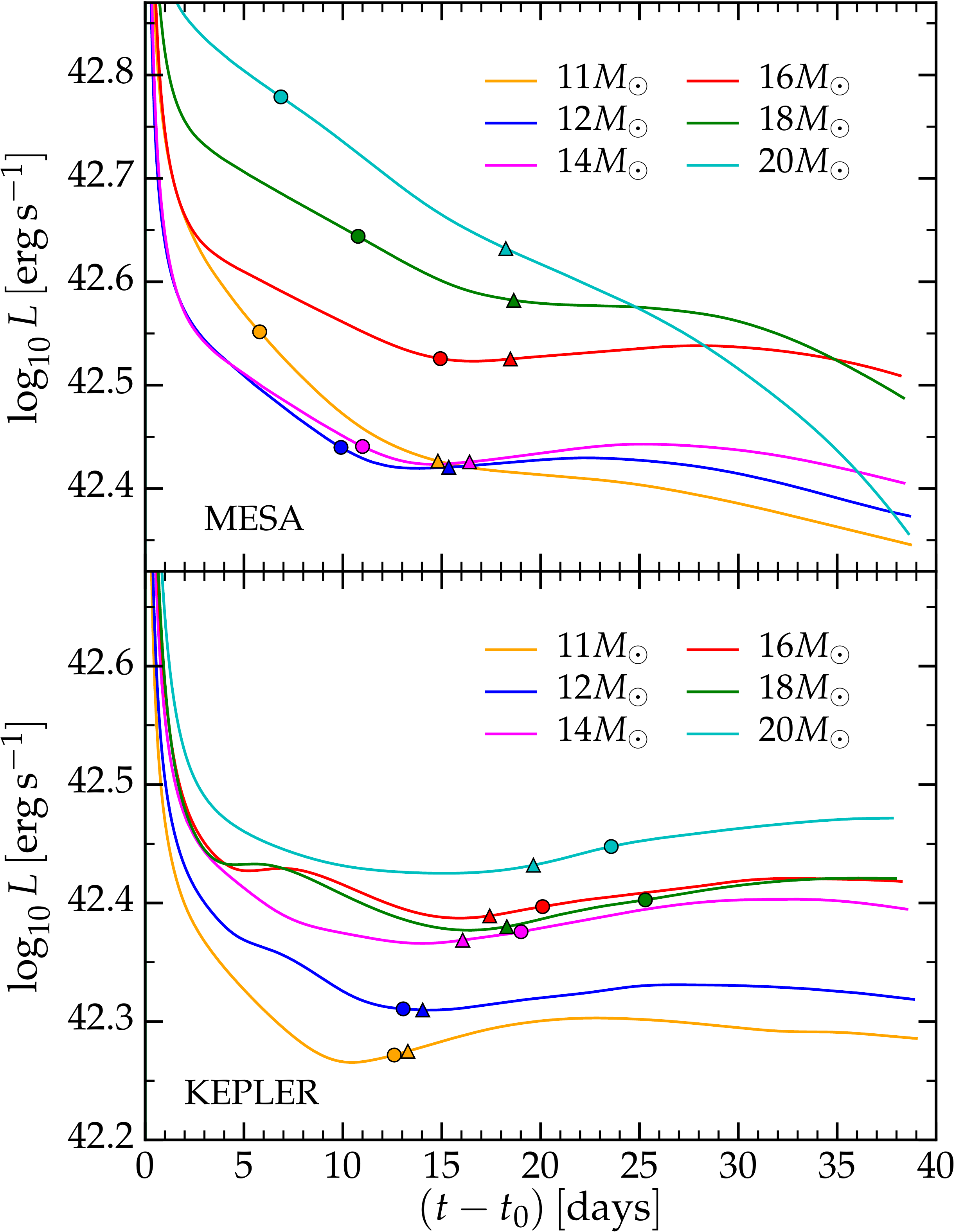}
  \caption{Bolometric light curves of the \texttt{MESA} (top panel)
    and \texttt{KEPLER} (bottom panel) models from
    Figure~\ref{fig:density} for a final explosion
      energy of $E_{\rm fin} = 10^{51}\,{\rm erg}$.  Colored circles
    indicate times when
    the luminosity shell (found via $\hat t_{\rm
      diff} = t - t_0$ and Equation~\ref{tdiff})
    enters the convective region of each envelope. Colored triangles
    roughly indicate when recombination begins
    (i.e. when effective temperature of the radiation
    drops down to $7500\,{\rm K}$).} \label{fig2a}
\end{figure}

The bottom panel of Figure~\ref{fig2} compares different ways for
calculating the bolometric luminosity of the same
model. The magenta solid curve shows the
bolometric luminosity at the photosphere ($\tau = 2/3$), as returned
by \texttt{SNEC}. Blue and green dashed curves show the light curves
computed as $\hat{E}/(t-t_{0})$ using the two different
discussed conditions for determining the location of the luminosity
shell.  To compute $\hat{E}$ we use
the \texttt{SNEC} output for the internal energy at each grid point
and find the total amount of this energy between the luminosity shell
and the photosphere.  Figure~\ref{fig2} shows that the bolometric
luminosity at the photosphere agrees well with the luminosity
calculated as $\hat{E}/(t-t_{0})$ until $\sim10-20$ days
after shock breakout, provided the condition $\hat
t_{\rm diff} = t - t_0$ using Equation~\ref{tdiff} is employed to find
the location of the luminosity shell.  At the same time, the
luminosity $\hat{E}/(t-t_{0})$ calculated from the condition $\hat\tau
= c/\hat{v}$ considerably underestimates the photospheric luminosity
and has a different slope.  This difference is because at larger
luminosity depth (see the top panel of Figure~\ref{fig2}) there is
more energy available and thus the shock cooling is more luminous when
the depth is calculated correctly.

We have already seen that at early times the light curves are
determined by a polytrope with $n\neq 1.5$, but what about at later
times when $n=1.5$? Whether or not this is ever satisfied in the shock
cooling phase depends on when recombination starts being important.
Figure~\ref{fig2a} shows the light curves of the models from
Figure~\ref{fig:density} for final energy $E_{\rm
  fin} = 10^{51}\,{\rm erg}$. Colored circles indicate the time
$t_{\rm conv}$ when the luminosity shell of each model computed as
$\hat t_{\rm diff} = t - t_0$ enters the convective part of the
model's envelope. Colored triangles indicate times when the effective
temperature of the radiation goes down to $7500\,{\rm K}$, which we
take as a rough criterion for the onset of recombination.  From
Figure~\ref{fig2a} it is clear that for most models the time it takes
for the luminosity shell to reach the convective part of the envelope
is comparable to the time when recombination sets in.  We therefore
conclude that there is rarely much time during which the emission from
shock cooling would be consistent with coming from stellar structure
described by an $n = 1.5$ polytrope (although a few of the \texttt{MESA} models
are exceptions in that the $n=1.5$ part of their
  envelope controls their cooling emission for $\sim 5-10$ days).

\begin{figure}[t]
  \centering
  \includegraphics[width=0.475\textwidth]{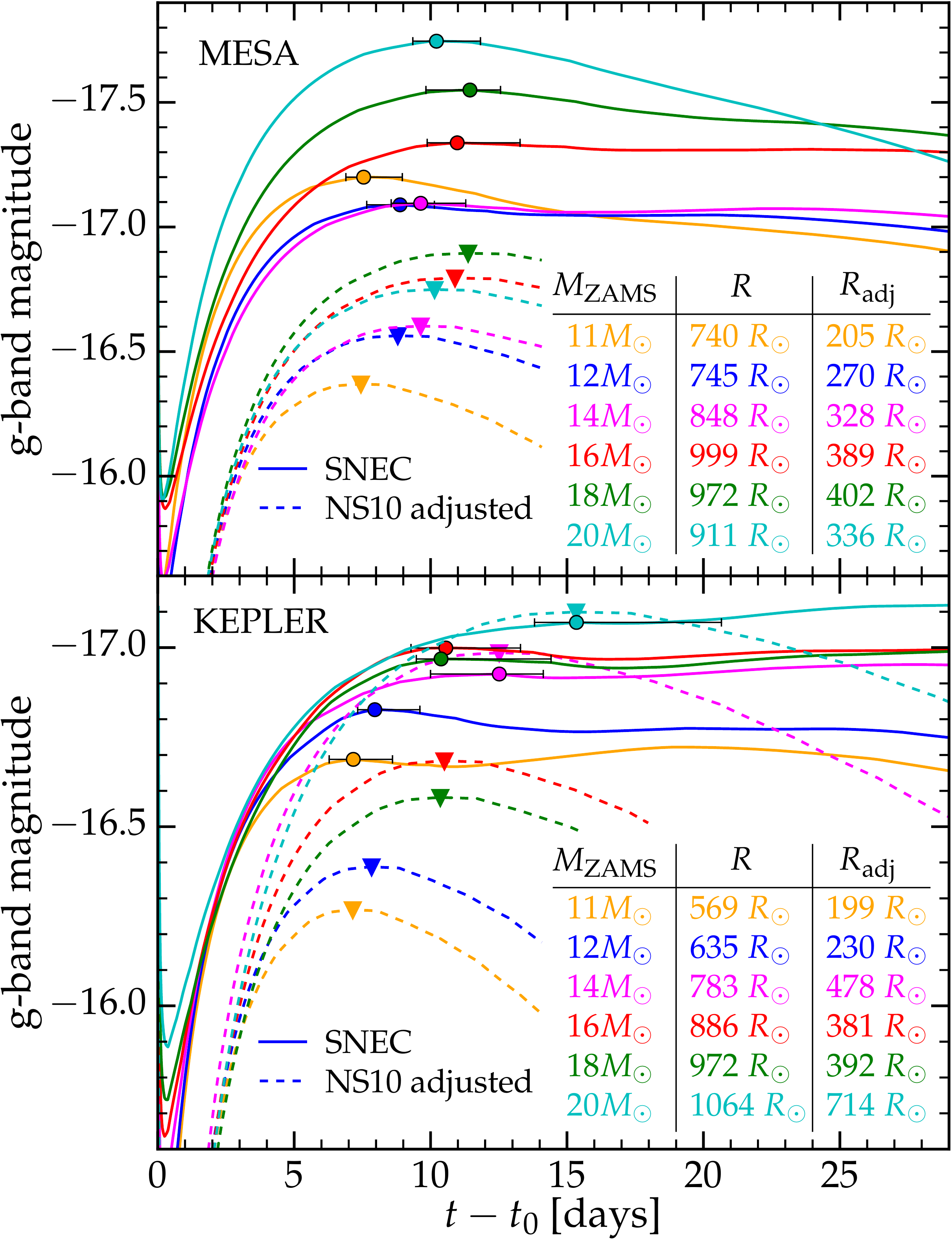}
  \caption{The light curves in $g$-band for the \texttt{MESA} (top
    panel) and \texttt{KEPLER} (bottom panel) models from
    Figure~\ref{fig:density} for $E_{\rm fin} = 10^{51}\,{\rm erg}$.
    Solid lines show the output of \texttt{SNEC}, and colored circles
    indicate their local maxima. Dashed lines are obtained from
    Equations (29) and (31) of NS10 by varying the radius in order to
    reproduce the rise times of the numerical light curves while
    keeping the mass and the final energy fixed. Colored triangles
    indicate the maxima of the analytic light curves, and $R_{\rm
      adj}$ gives the radius obtained in this way
    (cf.~\citealt{gonzalez-gaitan:15}).  The error bars show the
    time intervals during
    which the magnitude differs from the maximum by less than
    $0.01\,{\rm mag}$.} \label{fig3}
\end{figure}


\section{The rise times of SNe IIP}  
\label{sec2}

During the last several decades, numerical modeling of SN IIP
bolometric light curves was focused on reproducing their gross
properties, such as the length and the luminosity of the plateau
\citep[see][]{popov:93,kasen:09,bersten:11}. However, as was shown in the
analytical works of NS10 and \citet{goldfriend:14}, the slopes of the
light curves during the shock-cooling and plateau phases encode
important information on the structure of the density profiles of the
progenitor stars. Increasing abundance and quality of observational
data has made it possible to conduct systematic studies of the early
slopes for large sets of SN~IIP light curves
\citep{anderson:14,gonzalez-gaitan:15}.  Modeling the slopes of the
observed light curves and deducing the characteristics of the
progenitor stars based on these slopes will be a very important task
for future research.

In the current work, we instead focus on simply modeling the rise
times of SNe IIP. Being more robust and easier to measure than the
power law exponent of the bolometric light curve, the rise time can
still give us important information on the progenitor
characteristics. Our work is also
motivated by a number of recent works considering large sets of early
SN~IIP light curves and focusing on their rise times
\citep[e.g.,][]{gonzalez-gaitan:15,rubin:16,gall:15}.

\begin{figure}[t]
  \centering
  \includegraphics[width=0.475\textwidth]{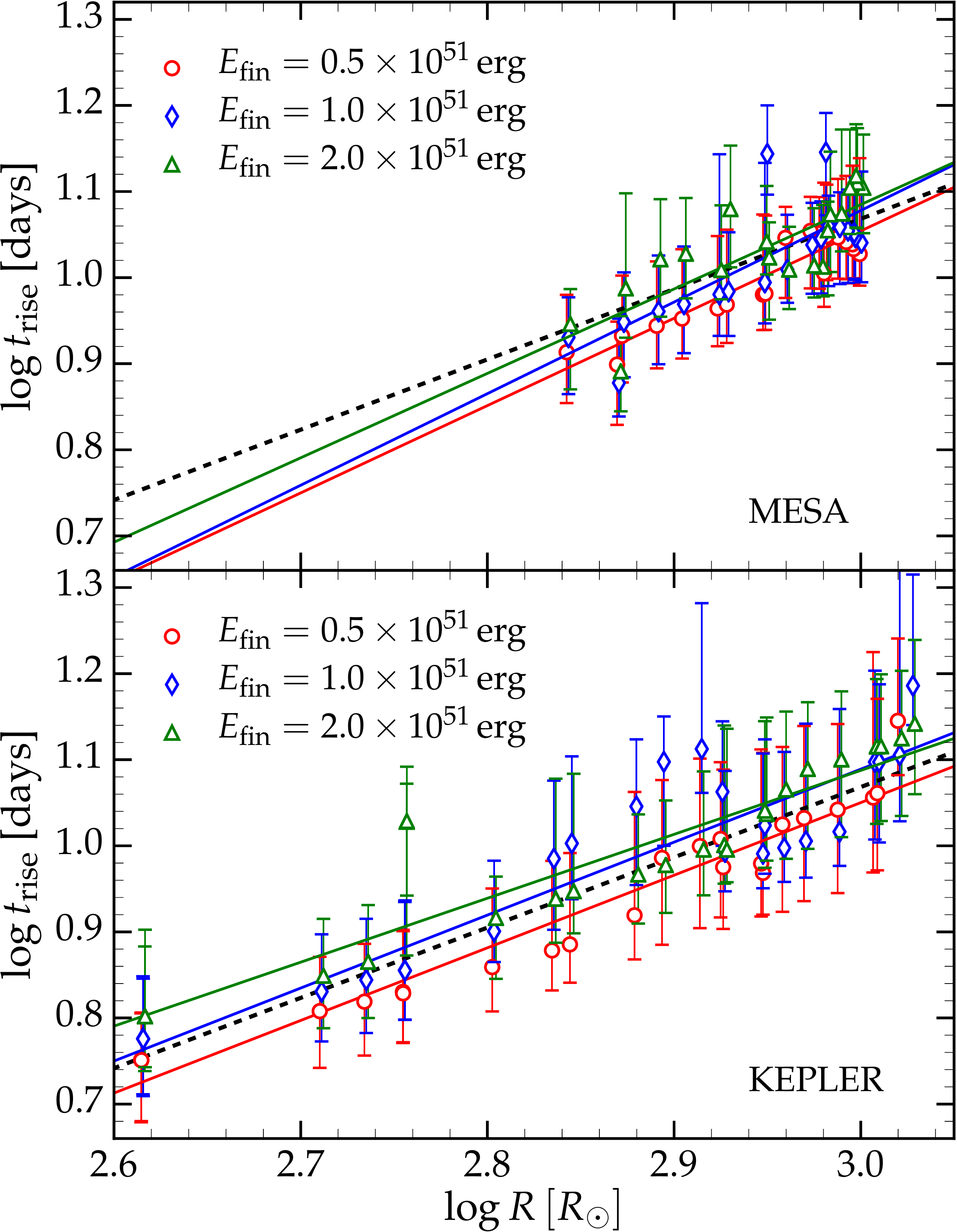}
  \caption{The g-band rise time $t_{\rm rise}$ as a function of the
    progenitor radius $R$ for \texttt{MESA} (top panel) and
    \texttt{KEPLER} (bottom panel) models using three different values
    for the final explosion energy. Error bars
    denote the time over which the $g$-band absolute magnitude lies
    within $0.01\,{\rm mag}$ of the maximum (see
    Figure~\ref{fig3}). Solid colored lines are linear fits to the
    same colored points in each panel. The black dashed line is the
    common fit for all models given by
    Equation~(\ref{fit}).} \label{fig4}
\end{figure}

\begin{figure}[t]
  \centering
  \includegraphics[width=0.475\textwidth]{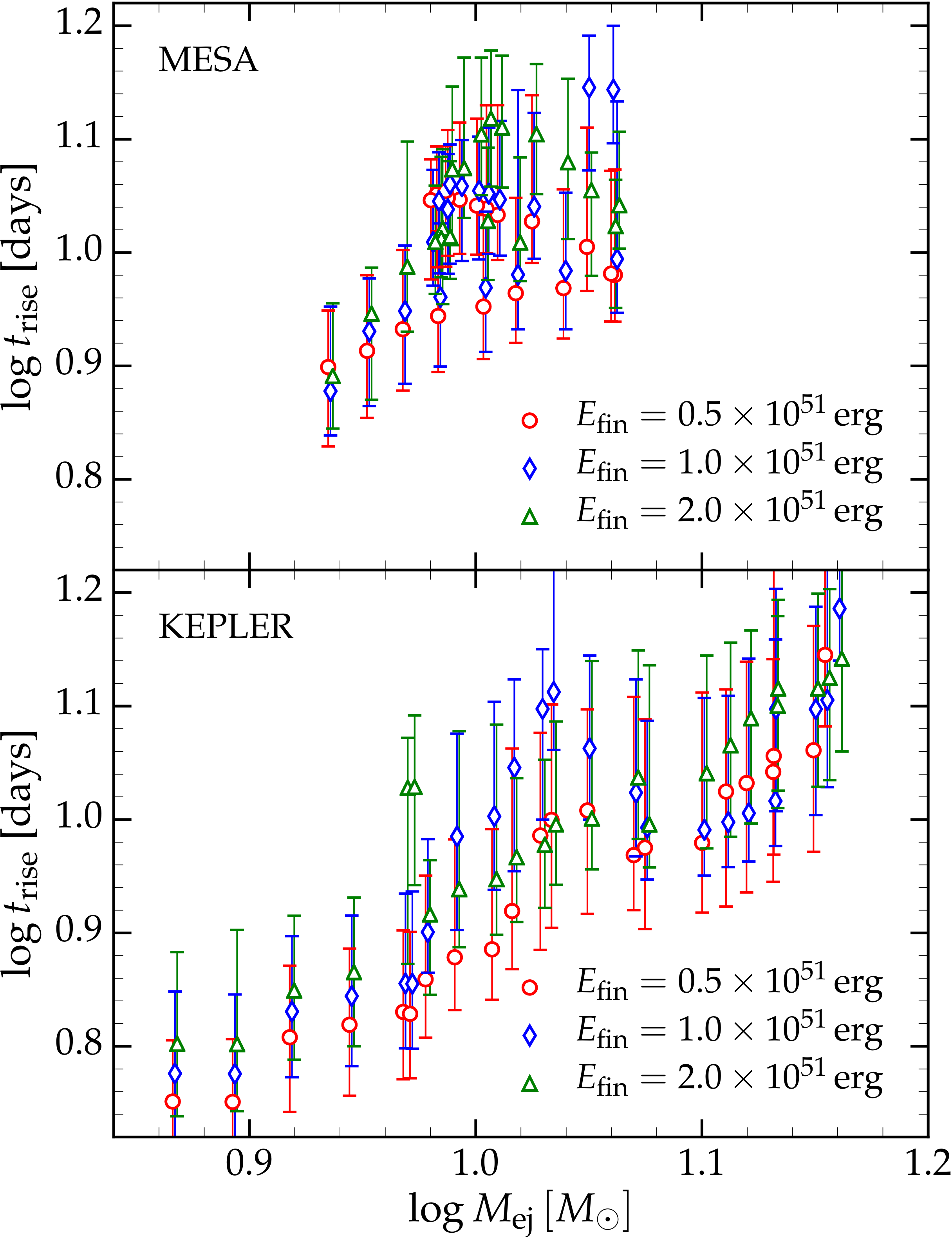}
  \caption{The g-band rise time $t_{\rm rise}$ as a function of the
    ejecta mass $M_{\rm ej}$ for \texttt{MESA} (top panel) and
    \texttt{KEPLER} (bottom panel) models at the three considered
    values of the final explosion
    energy.} \label{fig5}
\end{figure}

Figure~\ref{fig3} shows $g$-band light curves for a respresentative
set of models from Figure~\ref{fig:density}, for a final energy $E_{\rm fin}
= 10^{51}\,{\rm erg}$.  In our calculations, the SN emits as a blackbody 
at all times. Although the Rayleigh-Jeans part of the spectrum should 
not be strongly affected by this assumption during the shock cooling 
phase \citep{tominaga:11}, this will be more directly addressed in 
future work \citep[][in preparation]{shussman:16a}.
The light curves generated by \texttt{SNEC}
are plotted with solid lines. The colored circles indicate the local
maxima of the light curves. Here we define the rise time, $t_{\rm
  rise}$, as the time between shock breakout and the local maximum of
the color light curve in a given band.  The error bars show the
intervals of time during which the magnitude differs from the maximum
by less than $0.01\,{\rm mag}$. Later in this paper
(Figures~\ref{fig4} and~\ref{fig5}), we take this criterion as a
definition of the uncertainty with which the rise time can be
measured.  We choose $g$-band for the current study because in
$u$-band the light curves are more sensitive to non-thermal effects
like iron-group line blanketing \citep[see Figure 8 of][]{kasen:09},
which are not properly taken into account in \texttt{SNEC}.
Furthermore, $g$-band is a good match to many current and future
optical surveys.  In $r$-band, the light curves become very flat,
which makes it difficult to robustly identify the maximum.

We note that in some of our calculations we find
that the rise time depends on the duration of the thermal bomb.  To
investigate this dependence, we modeled the light curves of
the \texttt{KEPLER} models for bomb durations of
$0.1\,{\rm s}$, $10^{-2}\,{\rm s}$, $10^{-3}\,{\rm s}$, and
$5\times10^{-4}\,{\rm s}$. The vast majority of models
  show rise times independent of the bomb duration. The largest
  variations in $t_{\rm rise}$ ($\sim 4\,{\rm days}$) are seen for
 models with ZAMS masses in the range
between $\sim13\,M_{\odot}$ and $\sim16\,M_{\odot}$
and at $E_{\rm fin} = 2\times10^{51}\,{\rm
  erg}$. These models have fully converged rise times
  only at $t_\mathrm{bomb} \le 10^{-3}\,\mathrm{s}$.
 We believe that the observed
  dependence is due to the way the outer core structure in these
  models reacts to abrupt vs.\ to more gradual energy input. In
  first-principles core-collapse supernova simulations (e.g.,
  \citealt{bruenn:16}), the initial energy input by the expanding
  shock is abrupt, but most of the explosion energy builds up only
  gradually over $\sim$$1\,\mathrm{s}$. Since our thermal bomb
  approach does not allow us to fully realistically model the energy
  injection, we chose a thermal bomb duration of
  $10^{-3}\,\mathrm{s}$, which gives converged rise times for all our
  models.

For comparison with \texttt{SNEC} results, the dashed
lines in Figure~\ref{fig3} show the light curves obtained from
Equations (29) and (31) of NS10 for RSGs.  For the values of mass and
energy in these equations we use the ejecta mass $M_{\rm ej}$ and the
final energy $E_{\rm fin}$ of each model. We choose the radius
in these equations, $R_{\rm adj}$, in
such a way that the maximum of the analytical light curve coincides in
time with the maximum of the corresponding numerical \texttt{SNEC}
light curve (the colored triangles indicate the maxima of the
analytical light curves). By doing so we mimic the way in which the
progenitor radii were derived from the rise times in
\citet{gonzalez-gaitan:15}.  One can see that the values of $R_{\rm
  adj}$ are on average a factor of $\sim 1.5-2.5$ smaller than the
actual radii of the RSG models.  Taking this factor into account would
bring the results of \citet{gonzalez-gaitan:15} in much better
agreement with the observed radii of RSGs
\citep[see][]{levesque:05,levesque:06} and with the
  radii of the RSG models shown in Figure~\ref{fig:models}.

As an aside, we note that from Figure~\ref{fig3} it may seem that the
numerical and analytical light curves have different breakout times
(note the offset of the light curves at early times). This is
explained by the fact that during the planar phase of post-explosion
expansion (first $1-2$ days), the $g$-band flux of the NS10 model goes
down and starts to rise once the spherical phase of the expansion
begins \citep[this is also seen in the right hand panel of Figure 1
  of][]{gonzalez-gaitan:15}.

We next run a full collection of explosions using all stellar models
and final energies of $0.5\times10^{51}\,{\rm erg}$, $10^{51}\,{\rm
  erg}$, and $2\times10^{51}\,{\rm erg}$. The main results of this set
of calculations are summarized in Figures~\ref{fig4} and~\ref{fig5},
where we plot g-band $t_{\rm rise}$ for all considered models as a
function of $R$ and $M_{\rm ej}$, respectively. Figure~\ref{fig4}
demonstrates that there is a strong correlation between the rise time
and the radius of the progenitor. Furthermore, both the \texttt{MESA}
and \texttt{KEPLER} models show similar dependences on $R$. On
average, larger final energies have longer rise times, but this
dependence is less strong. On the other hand, when we compare the rise
time to the ejecta mass in Figure~\ref{fig5}, the results are more
mixed. The \texttt{KEPLER} models (bottom panel) show some
correlation, but the spread is much larger than in
the radius comparison. In addition, some of this dependence clearly
comes from the strong correlation between mass and radius in the
\texttt{KEPLER} models (Figure~\ref{fig:models}). The \texttt{MESA}
models (top panel) show even less correlation with different masses
sometimes having the same rise time.

In principle, one should be able to derive the dependence of the rise time on
the radius and ejecta mass of the model in the form
$\log t_{\rm rise} = a_1 \log M_{\rm ej} + a_2 \log R + a_3$ with constant
coefficients $a_1$, $a_2$ and $a_3$. Unfortunately, correlations between 
radius and mass as shown in Figure~\ref{fig:models} make it difficult to 
isolate these dependencies. As a consequence,  
the values of $t_{\rm rise}$ for our models
form nearly a line in the three-dimensional space 
$(\log M_{\rm ej}, \log R, \log t_{\rm rise})$, and we find that
it is not conclusive to fit a surface of the form
$\log t_{\rm rise} = a_1 \log M_{\rm ej} + a_2 \log R + a_3$.
In the future, this dependence could possibly be inferred
by fitting a much larger range of models with 
different masses and radii.

On the other hand, comparing the \texttt{MESA} models in the top panels of
Figures~\ref{fig4} and~\ref{fig5} reveals that there is a much clearer
mapping between radius and rise time than between ejecta mass and rise
time. This is especially apparent in the mass dependence of
the rise time in the \texttt{MESA} models (top panel
of Figure~\ref{fig5}), which shows that the same ejecta masses can
have different rise times when the radius is different. This was
generally expected from previous analytic work (NS10,
\citealt{piro:13}), but it is helpful how clearly it is seen
here. Furthermore, Figure~\ref{fig4} shows that both the \texttt{MESA}
and \texttt{KEPLER} models have similar relationships (both slope and
normalization) between radius and rise time.  Motivated by this, we
find a linear fit for $\log t_{\rm rise}$ as a function of $\log
R$. Colored lines in Figure~\ref{fig4} show the fits taken separately
for each set of models and each value of the final energy.  The black
dashed line is the common fit in g-band for all models and all final
(explosion) energies, which has the following numerical
coefficients,
\begin{equation}
\label{fit}
\log R\,[R_{\odot}] = 1.225 \log t_{\rm rise}\,[{\rm day}] + 1.692 \ .
\end{equation}
From Figure~\ref{fig4} one can see that the slope of the radius
dependence is fairly insensitive to the model type (\texttt{KEPLER} or
\texttt{MESA}) and final energy. Equation~(\ref{fit}) can be used as a
tool to infer the radii of SN~IIP progenitors from
 current and future transient surveys. This will be
especially useful for analyzing large collections of events where it
is impractical to do individual modeling.

An analogous relation to Equation~(\ref{fit}) between the rise time in
the optical part of the spectrum and the progenitor characteristics
was obtained in \citet{gall:15}. From the analytical model of
\citet{arnett:80,arnett:82}, they found that the optical rise time
scales as $t_{\rm opt-rise}\propto R^{1/2} T_{\rm peak}^{-2}$, while
for the \citet{rabinak:11} model, $t_{\rm opt-rise}\propto R^{0.55}
T_{\rm peak}^{-2.2} E_{51}^{0.06}M_{\rm ej}^{-0.12}$, where $T_{\rm
  peak}$ is the effective temperature at the peak and $E_{51}$ is the
 final explosion energy in units of $10^{51}\,{\rm
  erg}$. In comparison, we find a somewhat stronger relation of
$t_{\rm rise}\propto R^{0.816}$, although we do not
  include the temperature dependence. In the
future, by running a large set of models with more diversity in their
ejecta-mass--radius relations, we will be able to
  better study how $t_{\rm rise}$ depends on
other factors besides radius.

Assuming that the rise times we find are representative of what is
found in nature, we can also ask what should be expected for the rise
times in a larger sample of events. Since the distribution of massive
stars is understood in terms of an initial mass function (and not a
radius function), we have to assume a mass-radius relation to explore
this. In this case we use the \texttt{KEPLER} models and a final
energy $E_{\rm fin} = 10^{51}\,{\rm erg}$.  Using Monte Carlo
techniques, we generate a large sample of massive stars with ZAMS
masses in the range between $8\,M_{\odot}$ and $20\,M_{\odot}$
(motivated by the masses approximately inferred by pre-explosion
imaging of SNe~IIP, \citealt{smartt:09}) with a Salpeter initial mass
function of
\begin{equation}
\frac{dN}{dM_{\rm ZAMS}}\propto M_{\rm ZAMS}^{-2.35}\ ,
\end{equation}
and calculate their rise times. The median value of the derived set of
rise times is $7.27$ days, which is
similar to the observed median value of $7.5\pm 0.3$ days
found by \citet{gonzalez-gaitan:15}. This
implies a median radius of $559\,R_{\odot}$
for SNe~IIP progenitors. This value is interesting for comparison to observational
samples of massive stars \citep[e.g.,][]{levesque:05,levesque:06}.


\section{Conclusions and discussion}
\label{sec3}

Using \texttt{SNEC} \citep{morozova:15}, we have exploded a set
of massive star models from both the \texttt{KEPLER} and \texttt{MESA}
stellar evolution codes to study how the rise of the SN
light curve depends on 
progenitor and explosion characteristics. We find that the strongest
correlation is between the g-band rise time and the radius at the time
of explosion, and provide a formula that relates these two properties
in Equation~(\ref{fit}). This can be used in future
analyses of SNe~IIP observations.

To better understand what properties of the progenitor are controlling
the early light curve, we examined the envelopes of red supergiants
obtained with both stellar evolution codes.  We find that their
convective envelopes do not extend all the way to the surface.  In
fact, all the considered models have regions close to their surface
where the power-law exponent $n$ is smaller than $1.5$ and its value
varies between different ZAMS
masses. These regions are important for the early light curve, since
the luminosity shell passes through them during the first $10-25$ days
after shock breakout. Due to the shallow density profiles in these
regions, the simple estimate $\hat{\tau} = c/\hat{v}$
  for the optical depth at the luminosity shell is inadequate. This
  explains the differences we find with previous semi-analytic
  treatments of the early light curve.

The results obtained in this paper add to a long standing discussion
about the progenitor radii of SNe IIP.  Based on the results of their
numerical models, \citet{dessart:13} showed that the color
properties of SNe IIP may be explained by small progenitor models
($\sim 500\,R_{\odot}$), while larger progenitors would produce light
curves that remain too blue for too long. \citet{gonzalez-gaitan:15}
came to a similar conclusion based on a comparison of the observed
rise times to analytical models of the shock-cooling phase.  The radii
they obtain are a factor of $\sim 2$ smaller than the observed radii
of RSGs \citep[see][]{levesque:05,levesque:06}, and have the median
value $\sim 350\,R_{\odot}$.  Progenitor radii $\lesssim
500\,R_{\odot}$ are also suggested by the results of
\citet{shussman:16} and \citet{garnavich:16}.  At the same time, the works of
\citet{valenti:14} and \citet{bose:15} for two SNe IIP deduce large
radii (up to $\sim 800\,R_{\odot}$) for their progenitor
stars. \citet{rubin:16} find that the progenitor radii are weakly
constrained by comparison to analytical shock-cooling models (at least
based on $R$-band photometry alone).  In this paper, we have shown
that deriving the progenitor radius in the way done in
\citet{gonzalez-gaitan:15} may considerably underestimate it. On the
other hand, exploding the \texttt{MESA} and \texttt{KEPLER} stellar
evolution models without reducing their radii, we find a reasonable
agreement with the median value of the observed $g$-band rise times
from \citet{gonzalez-gaitan:15}.

One of the main strengths of our technique is that it does not require
that the stellar models we use exactly replicate the progenitors that
exist in nature. Instead, having a more diverse sample of models with
different mass-radius relations gives us an increasingly better handle
on how the early rise depends on the progenitor properties.
Therefore, for future work it will be useful to explode an even larger
set of stellar models. This would help better test the relation we
find between rise time and radius, but also help us
  to better understand how sensitive the rise time is to the ejecta
mass. In addition, one could explore whether $T_{\rm peak}$, the
temperature at peak, is useful for tightening this relation, as has
been found in previous semi-analytic work \citep[see][]{gall:15}. In
this way, a fairly easy additional observable, namely the colors at
peak, could be used to make these radius measurements more
robust. This will be useful as a tool for future transient surveys, as
well as for comparison with pre-explosion imaging of SNe and studies of
massive stars that hope to connect these progenitors to the SNe they
will eventually make.

\acknowledgments

We acknowledge helpful discussions with and feedback from D.~Clausen,
L.~Dessart, R.~J.~Foley, S.~R.~Kulkarni, K.~Maeda, O.~Pejcha, R.~Sari,
D.~Radice, B.~J.~Shappee, and T.~Sukhbold.  We thank T.~Sukhbold for
providing us with the pre-collapse stellar evolution models from
\texttt{KEPLER}. We thank Ehud Nakar for helpful feedback on our paper 
and also for sharing a nearly complete draft of his own work (Shussman et al., 
in preparation).  \texttt{SNEC} and the light curves presented here
are available from \url{https://stellarcollapse.org/Morozova2016}.
This work is supported in part by the National Science Foundation
under award Nos.\ AST-1205732 and AST-1212170, by Caltech, by the
Sherman Fairchild Foundation, and by the International Research Unit
of Advanced Future Studies, Kyoto University. The computations were
performed on the Caltech compute cluster Zwicky (NSF MRI-R2 award
no.\ PHY-0960291), on the NSF XSEDE network under allocation
TG-PHY100033, and on NSF/NCSA Blue Waters under NSF PRAC award
no.\ ACI-1440083. This paper has been assigned Yukawa Institute for
Theoretical Physics report number YITP-16-32.

\bibliographystyle{apj}

\end{document}